\documentclass[aps,twocolumn,floats,prl,nofootinbib,superscriptaddress]{revtex4-1}

\usepackage{amsmath,amsfonts,amssymb,aas_macros,slashed}
\usepackage{bbold,wasysym}

\usepackage[usenames]{xcolor} 
\usepackage{soul}
\usepackage{physics}

\renewcommand{\sec}{\ensuremath{\mathrm{s}}}

\newcommand{\kg}{\ensuremath{\mathrm{kg}}}

\newcommand{\fm}{\ensuremath{\mathrm{fm}}}
\newcommand{\cm}{\ensuremath{\mathrm{cm}}}
\newcommand{\km}{\ensuremath{\mathrm{km}}}

\newcommand{\eV}{\ensuremath{\mathrm{eV}}}
\newcommand{\keV}{\ensuremath{\mathrm{keV}}}
\newcommand{\MeV}{\ensuremath{\mathrm{MeV}}}
\newcommand{\GeV}{\ensuremath{\mathrm{GeV}}}

\renewcommand{\day}{\ensuremath {\mathrm{day}}}

\let\oldhat\hat
\renewcommand{\vec}[1]{\mathbf{#1}}
\renewcommand{\hat}[1]{\oldhat{\mathbf{#1}}}

\usepackage{pgfplots}
\usepackage{xparse}

\usetikzlibrary{positioning,arrows,patterns}
\usetikzlibrary{decorations.markings}
\usetikzlibrary{calc}

\tikzset{
    photon/.style={decorate, decoration={snake}, draw=black},
    vector/.style={decorate, decoration={snake}, draw},
	provector/.style={decorate, decoration={snake,amplitude=2.5pt}, draw},
	antivector/.style={decorate, decoration={snake,amplitude=-2.5pt}, draw},
    fermion/.style={draw=black, postaction={decorate},
        decoration={markings,mark=at position .55 with {\arrow[draw=black]{>}}}},
    fermionbar/.style={draw=black, postaction={decorate},
        decoration={markings,mark=at position .55 with {\arrow[draw=black]{<}}}},
    fermionnoarrow/.style={draw=black},
    gluon/.style={decorate, draw=black,
        decoration={coil,amplitude=4pt, segment length=5pt}},
    scalar/.style={dashed,draw=black, postaction={decorate},
        decoration={markings,mark=at position .55 with {\arrow[draw=black]{>}}}},
    scalarbar/.style={dashed,draw=black, postaction={decorate},
        decoration={markings,mark=at position .55 with {\arrow[draw=black]{<}}}},
    scalartwo/.style={dotted,draw=black, postaction={decorate},
        decoration={markings,mark=at position .55 with {\arrow[draw=black]{>}}}},
    scalartwobar/.style={dotted,draw=black, postaction={decorate},
        decoration={markings,mark=at position .55 with {\arrow[draw=black]{<}}}},
    scalarnoarrow/.style={dashed,draw=black},
    electron/.style={draw=black, postaction={decorate},
        decoration={markings,mark=at position .55 with {\arrow[draw=black]{>}}}},
	bigvector/.style={decorate, decoration={snake,amplitude=4pt}, draw},
    vertex/.style={draw,shape=circle,fill=black,minimum size=1pt,inner sep=0pt},
}

\begin{document}

\title{Probing sub-GeV Dark Matter with conventional detectors}

\author{Chris Kouvaris}\thanks{kouvaris@cp3.sdu.dk}\affiliation{CP$^3$-Origins, University of
  Southern Denmark, Campusvej 55, DK-5230 Odense, Denmark}
\author{Josef Pradler}\thanks{josef.pradler@oeaw.ac.at}\affiliation{Institute of High Energy Physics,
  Austrian Academy of Sciences, Nikolsdorfergasse 18, 1050 Vienna,
  Austria}

\begin{abstract}
  The direct detection of Dark Matter particles with mass below the
  GeV-scale is hampered by soft nuclear recoil energies and finite
  detector thresholds. For a given maximum relative velocity, the
  kinematics of elastic Dark Matter nucleus scattering sets a
  principal limit on detectability.  Here we propose to bypass the
  kinematic limitations by considering the inelastic channel of photon
  emission from Bremsstrahlung in the nuclear recoil.  Our proposed method allows to set
  the first limits on 
  Dark Matter below $500\,\MeV$ in the plane of Dark
  Matter mass and cross section with nucleons. In situations where a
  Dark Matter-electron coupling is suppressed, Bremsstrahlung may
  constitute the only path to probe low-mass Dark Matter awaiting new
  detector technologies with lowered recoil energy thresholds.
\end{abstract}

\maketitle

{\em Introduction.}
Weakly interacting massive particles (WIMPs) are among the
theoretically best motivated and experimentally most sought particle
candidates for Dark Matter (DM)~\cite{Jungman:1995df,Bertone:2004pz}.
The efforts are driven by a broad expectation that physics beyond the
Standard Model (SM) should enter near the electroweak scale, with
interactions that are not too different from the weak interactions.

There has been a significant amount of experimental effort to push the
sensitivity of direct detection experiments to masses below a
few~GeV. The efforts are hampered by the fact that light DM induces
soft nuclear recoils that are difficult to detect unambiguously. In
the non-relativistic scattering of a DM particle~$\chi$ and target
nucleus~$N$ with mass $m_N$, the three-momentum transfer
$\vec q = \vec p_{\chi}' - \vec p_{\chi} $ 
determines the kinetic recoil energy of the nucleus,
$ E_R = {|\vec{q}|^2}/({2 m_N}) \leq {2 \mu_N^2 v^2}/{m_N}$,
where $\mu_N$ is the DM-nucleus reduced mass and $v$ is the relative
velocity, bounded by the finite gravitational potential of the
galaxy.
New avenues have therefore been suggested to probe DM below the
GeV-scale, such as looking for DM-electron
scattering \cite{Essig:2011nj} in existing data~\cite{Essig:2012yx}, employing semiconductor
targets~\cite{Graham:2012su,Essig:2015cda,Lee:2015qva}, using
superconductors or superfluids~\cite{Guo:2013dt,Schutz:2016tid,
  Hochberg:2015pha, Hochberg:2015fth}, nanotubes~\cite{Cavoto:2016lqo}, 2D graphene-like targets~\cite{Hochberg:2016ntt} and exploiting a
non-virialized velocity-component of DM~\cite{Kouvaris:2015nsa}.

In this letter we propose a method of probing sub-GeV DM in direct
detection by going to the inelastic channel of photon emission from
the nucleus in form of Bremsstrahlung%
\footnote{Photon emission from the excitation of low-lying nuclear
    levels has been considered
    in~\cite{Ellis:1988nb,Baudis:2013bba,McCabe:2015eia,Vergados:2016ytt}. The process
    requires considerable momentum transfer and concerns electroweak
    scale DM masses.}%
---an irreducible
contribution that accompanies the elastic reaction,
\begin{subequations}
\label{eq:processes}
  \begin{align}
  \label{eq:elastic}   \chi + N &\to \chi + N(E_R) \,  && \text{(elastic)},\\
  \label{eq:brems}  \chi + N &\to \chi + N(E_R') + \gamma(\omega)\, && \text{(inelastic)}. 
  \end{align}
\end{subequations} 
The virtue of considering (\ref{eq:brems}) is that the available
photon energy is bounded by the energy of the relative motion of DM
and the target, $\omega \leq \mu_N v^2 /2 $, so that we observe a
hierarchy for light dark matter,
\begin{align}
E_{R, \rm max} = 4(m_{\chi}/m_N)\omega_{\rm max} \ll \omega_{\rm max} \quad (m_{\chi}\ll m_N). 
\end{align}
As we will see, the larger energy deposition in photon emission allows
to lower the sensitivity to nuclear recoils to the sub-GeV DM mass
regime in present-day detectors.  The signal will be be part of the
``electron recoil-band'' and subject to backgrounds, yet amply
detectable: whereas, say, $E_R = 0.5\,\keV$, is experimentally easily
missed, a photon of energy $\omega = 0.5\,\keV$ is hardly ever missed.

\begin{figure}[tb]
  \centering
\includegraphics[width=\columnwidth]{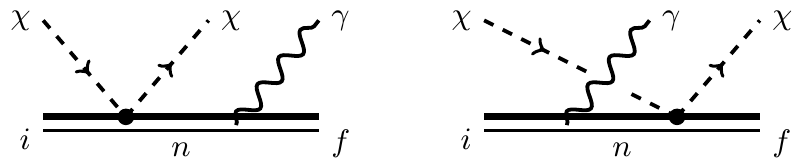}%
\caption{\small Photon emission resulting  from DM-nucleus
  scattering. The thick line represents the nucleus in the atomic
  initial (final) state $i$ ($f$) with intermediate state $n$,
  represented by the thin solid line.}
  \label{fig:bremsDD}
\end{figure}

\paragraph{Cross section.}

It is well known that in the limit of soft photon emission off an electromagnetically
charged particle the matrix element for Bremsstrahlung factorizes into
the matrix element of elastic scattering $M_{\rm el }$ times a
manifestly gauge invariant piece.
For this to hold in the non-relativistic limit of the emitting
particle, the three momentum transfer $\vec q$ in the elastic
scattering must be much larger than the change of it due to the
additional emission of the photon with momentum $\vec k$,
$  \delta \vec q = (\vec p_N' - \vec p_N - \vec k) - (\vec p_N' - \vec p_N)_{\omega =0}  $. 
Hence, imposing $| \delta \vec q | \ll | \vec q | $ yields the
soft-photon limit,
$  \omega \ll |\vec q| v = \sqrt{2 m_N E_R} v  \simeq  O(10\,\keV) \sqrt{\frac{A }{130}} \sqrt{\frac{E_R}{1\,\keV}} $,
where $A$ is the atomic mass number of $N$. 
The latter condition holds well away from the kinematic endpoint of
minimum momentum transfer, $|\vec q_{\rm min} | \simeq \omega/v$, and
the soft photon limit will be respected. 
Since
$\omega_{\rm max} \simeq \mu_N v^2 /2 \lesssim | \vec q_{\rm max}| v $
holds parametrically, we can further take the approximation
$E_R' \simeq E_R$ in~(\ref{eq:brems}).

The factorization of the matrix element is universal and does not
depend on the spin of the nucleus. Summing over the photon
polarization, and assuming no directional sensitivity yields a double
differential cross section of,
\begin{align}
\label{eq:naive}
   \left. \frac{d^2\sigma}{dE_R d\omega} \right|_{\rm naive} & =  
  \frac{4 Z^2 \alpha}{3\pi}  \frac{1}{\omega} \frac{E_R}{m_N} \times 
 \frac{d\sigma}{dE_R}  \Theta( \omega_{\rm max} - \omega ) . %
\end{align}
Here, $d\sigma/dE_R $ is the WIMP-nucleus elastic scattering cross
section for (\ref{eq:elastic}), $Z$ is the atomic number.

The cross section for Bremsstrahlung emission off a recoiling
nucleus gets modified at low photon energies by the fact that it is in
a neutral bound state with electrons. The process of photon emission
can be viewed as in Fig.~\ref{fig:bremsDD} where the double line
represents the nucleus in the initial (final) atomic state of
electrons~$i$~($f$), with intermediate state~$n$.
The matrix element for the transition can be put into the following
form,
\begin{align}
\label{eq:ME}
  | V_{fi} |^2 = 2\pi\omega | M_{\rm el}|^2 & \left| \sum_{n \neq i,f} 
 \left[  \frac{ ( \vec d_{fn} \cdot \hat e^{*} ) 
\mel{n}{e^{-i \frac{m_e}{m_N} \vec q \cdot \sum_{\alpha} \vec r_{\alpha}}}{i} }{\omega_{ni} - \omega} \right. \right. \nonumber \\
&
\!\!\!\!\!\!\!\!\!\!\!\!\left. \left. 
+  \frac{ ( \vec d_{ni} \cdot \hat e^{*} ) 
\mel{f}{e^{-i \frac{m_e}{m_N} \vec q \cdot \sum_{\alpha} \vec r_{\alpha}}}{n } }{\omega_{ni} + \omega}
 \right] \right|^2 .
\end{align}
Here, $M_{\rm el}$ is the matrix element for the elastic DM-nucleus
collision, $\vec d_{kl} = e \sum_{\alpha} \vec r_{\alpha, kl} $ is the
atomic dipole moment (with sum over the positions $\vec r_{\alpha}$ of
all electrons with elementary charge $e$), $\hat e^{*}$ is the
polarization vector of the photon in three-dimensional transverse
gauge, and $\omega_{kl} = \omega_k - \omega_l$ is the atomic
transition frequency between states $\ket{k}$ and $\ket{l}$. The cross
section for photon emission will then be given by 
\begin{align}
  d\sigma =  \frac{|V_{fi}|^2}{|M_{\rm el}|^2} \frac{\omega^2 d\omega d\Omega_{\vec k}}{(2\pi)^3}  \times  d\sigma_{\rm el} . 
\end{align}

A few comments regarding (\ref{eq:ME}) are in order. First, the
factors $\vec d_{kl} \cdot \hat e^{*}$ are part of the dipole
transition element
$V^{(\gamma)}_{ kl} = - \vec d_{kl} \cdot \partial_t \vec A^{*}_{\vec
  e,\omega} $
with $\vec A = (2\pi/\omega)^{1/2} e^{-i\omega t}\, \hat e$ (we work
in unrationalized units of $e^2=\alpha$), responsible for the emission
of a photon of energy $\omega$.  Here, the spatial dependence entering
the photon wave function through
$|\vec k \cdot \vec x| \leq \omega R_{\rm Atom}$ has been neglected;
this is a good approximation, unless one considers the kinematic
photon endpoint and substitutes for $R_{\rm Atom} $ the entire atomic
radius, for which the product can become $O(1)$.
Second, the matrix elements
$\mel{k}{e^{-i \frac{m_e}{m_N} \vec q \cdot\sum_{\alpha} \vec
    r_{\alpha}}}{l} $
describe the motion of the electron-cloud relative to the nucleus with
velocity $|\vec v_N| = |\vec q|/m_N$ after the latter receives an
impulse $\vec q$ from DM. It is assumed that the kick is to good
approximation instantaneous,~\textit{i.e.}, the DM-nucleus interaction
time $\tau_{\chi} \sim R_N / v_{\chi}$ is smaller than the time it
takes electrons in orbit to adjust to the perturbation,
$\tau_{\alpha} \sim |\vec r_{\alpha}|/v_{\alpha} $.  Taking for the
nuclear radius $R_N = 1.3\,\fm A^{1/3}$, a typical DM velocity
$v_{\chi}= 10^{-3}$, and an inner shell electron with radius
$|\vec r_{\alpha}| = 1/(Z\alpha m_e)$ and velocity
$v_{\alpha } \sim Z \alpha$, we get
$ \tau_{\chi}/\tau_{\alpha} \simeq 10^{-4} A^{1/3} Z^2 $. Hence our
approximation is well justified for light elements; for heavier
targets such as xenon, the ratio can become $O(1)$, but only for the
innermost electrons. Going beyond the mentioned approximations
requires a dedicated atomic physics calculation, which is certainly
welcome but well beyond the scope of this paper. Finally, in the
denominators of (\ref{eq:ME}) we neglect any dependence on $E_R$ based on the fact that $E_R\ll\omega_{ni}$.

On similar grounds as for the dipole matrix element for photon
emission, we can make use of the dipole approximation in the boosted
matrix elements,
\begin{align}
\mel{k}{e^{-i \frac{m_e}{m_N} \vec q \cdot\sum_{\alpha} \vec
    r_{\alpha}}}{l} \simeq   \frac{-i}{e} \frac{m_e}{m_N} \,  \vec q \cdot \vec d_{kl}   \quad (k\neq l).
\end{align}
The limit is well justified, since
$ \frac{m_e}{m_N} \vec q \cdot \vec r_{\alpha} \ll 1 $ for all
practical purposes. This expansion brings about a major simplification
when we consider the special case $i = f$:
\begin{align}
  |V_{ii}|^2 =  \frac{4\pi\omega m_e^2}{\alpha} \frac{E_R}{m_N} | M_{\rm el} |^2  \times | \hat e_r^{*} \hat q_s \alpha_{rs}(\omega) |^2 . 
\end{align}
Here, $\alpha_{rs}(\omega)$ ($r,s$ are cartesian coordinates) denotes
the polarizability of an individual atom. In the limit of spherical
symmetry, which we will assume henceforth,
$\alpha_{rs}(\omega) = \alpha(\omega) \delta_{rs}$. The latter
function $\alpha(\omega)$ can be related to the atomic scattering
factors $f(\omega) = f_1(\omega) + i f_2(\omega)$ which are tabulated,
$  \alpha(\omega) = - \frac{\alpha}{m_e \omega^2} f(\omega) $. 
By taking the limit in which the atom stays in the ground state,
$i = f$, we neglect further contributions to the photon yield.  Our
derived limits must therefore be considered as conservative; we leave
more detailed calculations of the atomic processes as future work.

Taking the polarization sum, integrating over the photon directions
$d\Omega_{\vec k}$ and averaging over the direction $\hat q$ of the
momentum transfer, we arrive at the final result for the
photon-emission cross section,
\begin{align}
\label{eq:dsigdErdomg}
  \frac{d^2\sigma}{d\omega dE_R} & = \frac{4  \omega^3 }{3 \pi} 
  \frac{ E_R}{m_N } \frac{m_e^2|\alpha(\omega)|^2}{\alpha}  \times \frac{d\sigma}{dE_R}  \Theta( \omega_{\rm max} - \omega ) \nonumber \\ 
   & = \frac{4 \alpha}{3\pi \omega} \frac{E_R}{m_N }  |f(\omega)|^2 \times \frac{d\sigma}{dE_R}  \Theta( \omega_{\rm max} - \omega ).
\end{align}
A comparison with (\ref{eq:naive}) exposes nicely the atomic physics
modification to the na\"ive cross section of unscreened Bremsstrahlung
emission from the bare nucleus. At low photon energy, the process
weakens as $\omega^3$ as is typical for dipole emission (the dipole
created between the nucleus and electrons). At large energies,
$f_1 \to Z \gg f_2$, the atomic state becomes irrelevant, and
(\ref{eq:dsigdErdomg}) approaches~(\ref{eq:naive}).

\paragraph{Event rates.}

The main idea is to tap the electron recoil that is induced by
Bremsstrahlung, when (reliable) experimental sensitivity to nuclear
recoils fails at low recoil energy.
To arrive at a convenient expression for the differential event rate
we neglect the energy deposition $E_R$ since the respective maximum
energies fulfill
$E_{R, \rm max} \ll \omega_{\rm
  max}$,
and take the photon energy $\omega$ as the only detectable signal,
with rate
$d\sigma/d\omega = \int_{E_{R,\rm min}}^{E_{R,\rm max}} dE_R
\frac{d\sigma}{dE_R d\omega} $.
The boundaries of the recoil energy integration are found from 3-body
kinematics,
\begin{align*}
  E_{R,\rm max/min} & = \frac{\mu_N^2 v^2}{m_N} \left[ \left(1 -\frac{ \omega}{\mu_N v^2  }\right) \pm   \sqrt{1 - \frac{2 \omega}{\mu_N v^2} }    \right] . 
\end{align*}

Consider now a standard DM-nucleus recoil cross section,
$d\sigma/dE_R = \sigma^{SI}_0 m_N/(2\mu_N^2 v^2) F^2(|\vec q|) $ with
spin-independent DM-nucleus cross section
$\sigma_0^{SI} \simeq A^2 \sigma_n (\mu_N/\mu_n)^2 $ where $\sigma_n$
is the DM-nucleon elastic cross section and $\mu_n$ the DM-nucleon reduced
mass. Making the excellent approximation that the nuclear form factor at
low recoil is unity, $F^2 \simeq 1$, the differential cross section can be
integrated to yield,
\begin{align}
\label{eq:dsigdomg}
  \frac{d\sigma}{d\omega} = 
 \frac{4\alpha |f(\omega)|^2}{3\pi \omega} \frac{\mu_N^2 v^2\sigma_0^{SI} }{m_N^2} 
  \sqrt{1-\frac{2\omega}{\mu_N v^2} } \left( 1 - \frac{\omega}{\mu_N v^2}   \right). 
\end{align}

In a final step, we take the average of the cross section over the
velocity distribution of DM in the frame of the detector and compute
the event rate,
\begin{align}
\label{eq:velavg}
  \frac{dR}{d\omega} = N_T \frac{\rho_{\chi}}{m_{\chi}} \int_{|\vec v|\geq v_{\rm min}}
  d^3\vec v\, v f_v(\vec v + \vec v_e)  \frac{d\sigma}{d\omega} .
\end{align}
Here, $N_T$ is the number of target nuclei per unit detector mass and
$\rho_{\chi} = 0.3 \,\GeV/\cm^3$ is the local DM mass density.  For
$f_v(\vec v)$ we take a truncated Maxwellian with escape speed
$v_{\rm esc } = 544\,\km/\sec$~\cite{Smith:2006ym} and most probable
velocity $v_0 = 220\,\km/\sec$; $\vec v_{e}$ is the velocity of the
Earth relative to the galactic rest frame and
$v_{\rm min} = \sqrt{2\omega/\mu_N}$.

\begin{figure}
\centering
\includegraphics[width=\columnwidth]{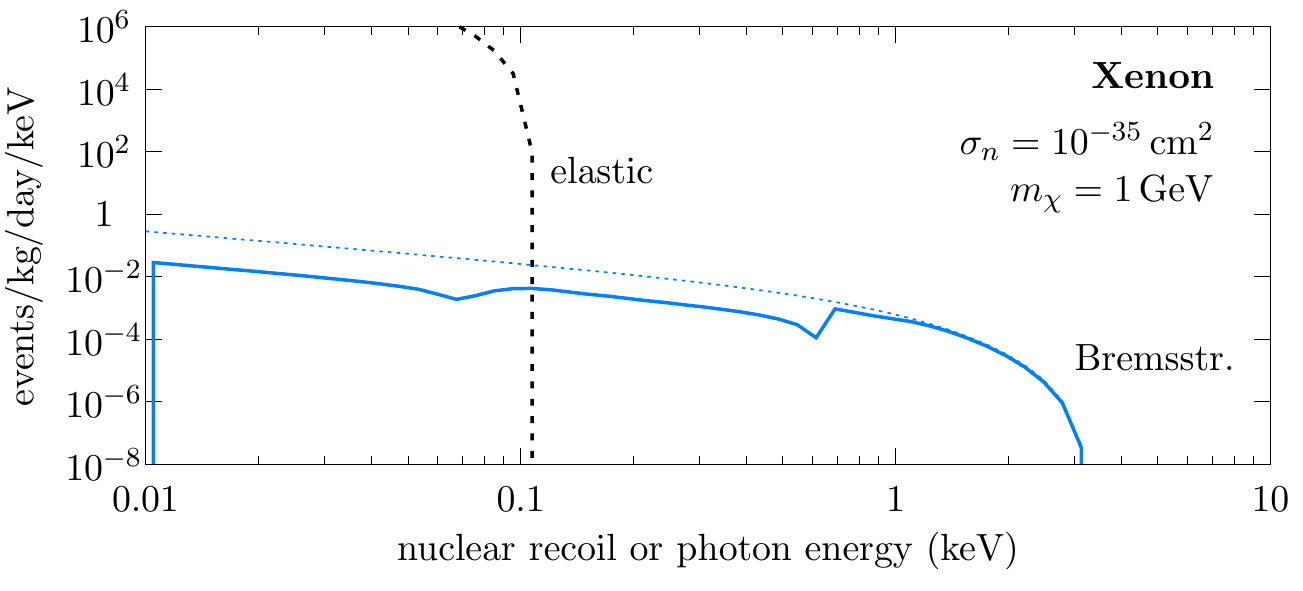}%
\caption{Elastic ($dR/dE_R$) and photon-emission $(dR/d\omega)$ rates
  in xenon. The ionization threshold is $12\,\eV$. The dotted line is
  derived from the na\"ive cross section~(\ref{eq:naive}).}
\label{fig:cs}
\end{figure}

The penalty for going to the inelastic channel is of course very
large.
Whereas a factor of $\alpha$ is compensated by $Z^2$ in
~(\ref{eq:naive}) [or by $f_{1,2}^2$ in (\ref{eq:dsigdomg})], the
factor $E_R/m_N$ may be overcome by a quasi-exponential rising event
rate $ dR_{\rm el}/dE_R \sim e^{-E_R/E_0}$ with decreasing $E_R$ where
$E_0 = {\rm few} \times \keV$ for WIMPs and typical target masses. The
spill over from photons into the higher energy region is the key that
allows us to exploit the inelastic channel in the electron recoil band
experimentally.

The prospective parameter space where the method of Bremsstrahlung
emission yields an improvement of sensitivity, is best identified by
demanding that no elastic nuclear recoil event (with rate $dR/dE_R$)
has been induced above the detector-specific nominal threshold recoil
energy $E_{R,\rm th}$,
$ N(E_R > E_{R,\rm th}) =  {\rm exposure} \times \int_{E_{R,\rm th}}^{\infty} dE_R \frac{dR}{dE_R} < 1 $,
and by computing from there the number of bremsstrahlung-induced electron
recoil events via~(\ref{eq:velavg}).
It is important to note that $ N(E_R > E_{R,\rm th}) < 1 $ becomes
trivially fulfilled for any value of DM-nucleon cross section once the
DM mass falls below the kinematic threshold imposed by the maximum relative
velocity between DM and target nucleus,
$v_{\rm max} = v_{\rm esc} + v_{e} \simeq 750\,\km/\sec$.
For example, $N(E_R > E_{R,\rm th}) < 1$ for any value of
$\sigma_0^{SI}$ once the DM mass falls below $3.3\,\GeV$ in a xenon
experiment with nominal threshold of $E_{R, \rm th} =1.1\,\keV$ such
as in LUX~\cite{Akerib:2013tjd,Akerib:2015rjg} and before accounting
for finite detector resolution.
Figure~\ref{fig:cs} shows the theoretical rates for elastic
scattering, $dR/dE_R$, and the photon emission rate $dR/d\omega$ as
labeled resultant from nuclear recoils of a DM particle of mass
$m_{\chi} = 1\,\GeV$ and a DM-nucleon cross section of
$\sigma_n = 10^{-35}\,\cm^2$. The dotted line is the rate according to
the naive estimate~(\ref{eq:naive}).

\paragraph{Probing low-mass DM.}

\begin{figure}
\centering
\includegraphics[width=\columnwidth]{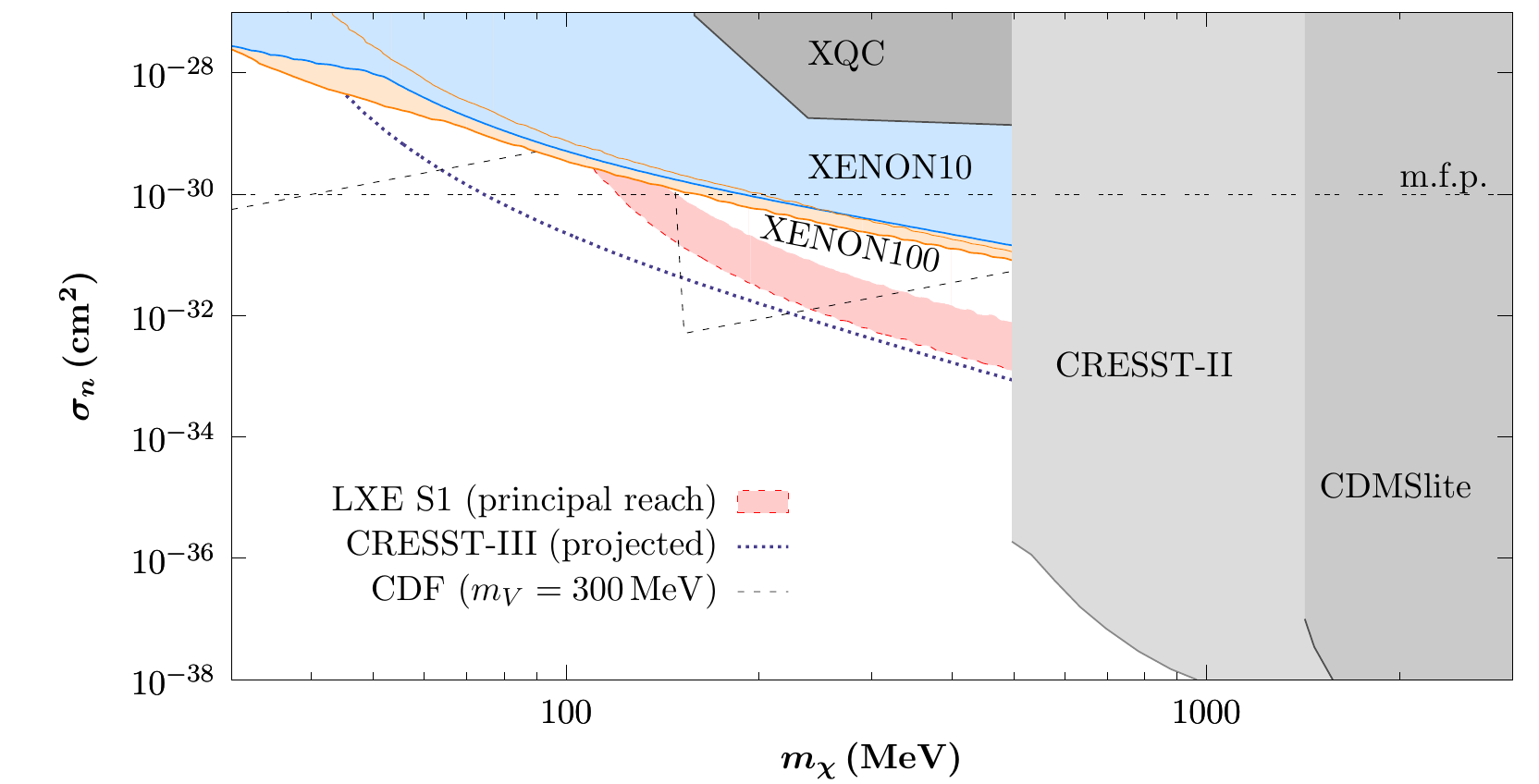}%
\caption{$(m_{\chi}, \sigma_n)$ plane of DM mass and DM-nucleon cross
  section. Regions labeled CRESST-II and CDMSlite were previously
  excluded from elastic DM-nucleus
  scattering~\cite{Angloher:2015ewa,Agnese:2015nto}; regions labeled
  XENON10 and XENON100 show newly derived constraints based
  on~(\ref{eq:brems}). Above the line ``m.f.p.''~limits are
  invalidated as DM scatters before reaching the detector. The region
  XQC is excluded from rocket-based X-ray calorimetry
  data~\cite{Erickcek:2007jv}. Projections for CRESST-III and for a
  dedicated liquid Xe experiment, labeled LXE, are also shown; see main text.  The
  monojet constraint CDF is model dependent; CMB constraints can be
  evaded~\cite{Batell:2014yra} and are not shown.}
\label{fig:summary}
\end{figure}

We now explore the sensitivity to Bremsstrahlung in the usual
$(m_{\chi},\sigma_n)$ plane. Here we focus on the ionization-only
signal in liquid scintillator experiments, for which
XENON10~\cite{Angle:2011th}, and most recently
XENON100~\cite{Aprile:2016wwo}, have presented results.  The
ionization threshold of xenon is $\sim 12\,\eV$, hence the emission of
a 100~eV photon can already produce multiple ionized electrons.

For XENON10 the collaboration has reported the spectrum in number of
electrons, and we compute the electron yield upon absorption of the
Bremsstrahlung photon following~\cite{Essig:2012yx,Bloch:2016sjj} and
assuming that it takes on average 13.8~eV to produce an ionized
electron~\cite{Shutt:2006ed}; a simliar, albeit simplified program
has been carried out in~\cite{An:2013yua,An:2014twa}.
For XENON100 we convert the expected ionization signal into
photoelectrons (PE) using a yield of $19.7~{\rm PE}/e^-$ and a width
of $7~{\rm PE}/e^-$~\cite{Aprile:2016wwo}; the conversion corresponds
to $1.43\,~{\rm PE}/\eV$.
Although signal formation at lowest energies is
poorly understood, a recent measurement at $200\,\eV$ electron recoil
energy supports such naive expectations of charge yield, with
recombination of ions and electric field dependence playing little
role~\cite{LUXtalk,Akerib:2015wdi}. We then place a limit using the
``$p_{\rm max}$'' method~\cite{Yellin:2002xd,Angle:2011th}.
The respective sensitivities to $\sigma_n$ are shown in
Fig.~\ref{fig:summary} by the (blue and orange) shaded region as
labeled. A thin orange line in the XENON100 region shows the limit
with the (ad-hoc) pessimistic choice of 30~eV/electron and resulting
conversion factor $0.6\,~{\rm PE}/\eV$. Finally, we note that LUX may soon
improve on the XENON100 limit, because of lower electron
backgrounds~\cite{McKinnseyprivate}.
We estimate (supported by our own Monte Carlo simulation; see also
  \cite{Collar:1992qc,Collar:1993ss,Hasenbalg:1997hs,Zaharijas:2004jv})
that for $\sigma_n \gtrsim 10^{-30}\,\cm^2$ the limits become
invalidated as elastic scattering of DM inside the Earth slowly
degrades energy and flux of the incident particles; the corresponding
approximate demarcation line is labeled ``m.f.p.''.

Importantly, the next generation of dual-phase liquid scintillator
direct detection experiments, such as XENON1T~\cite{Aprile:2015uzo}
and LZ~\cite{Malling:2011va}, are coming online or are being planned.
Although a much reduced electromagnetic recoil background may be
expected,
$R_{\rm b.g.}  =
10^{-4}-10^{-5}\,/\kg/\day/\keV$~\cite{Aprile:2015uzo},
such rate requires volume fiducialization. The latter is
  accomplished through the much weaker scintillation signal S1 with
an overall detection efficiency of only $\sim 10\%$.  Hence,
improvement over current limits is not guaranteed, and instead we
advocate a dedicated smaller setup with both, high S1 light collection
efficiency and single photon sensitivity~\cite{MattPyleprivate}; see
also recent Ref.~\cite{Derenzo:2016fse}. We estimate a principle limit
for this technology in Fig.~\ref{fig:summary}, assuming
$R_{\rm b.g.} = 10^{-4}\,/\kg/\day/\keV$ with an exposure of
$600$~kg-yr and requiring two S1 photons produced at 10\% with respect to
ionized electrons and each of the S1 photons detected with
  40\%-100\% efficiency (varying the efficiency determines the
thickness of the red band).
Finally, solid state scintillators have reached $O(100\,\eV)$
thresholds, most notably CRESST-II with the only reported DM-nucleon
cross section limit below~$1~\GeV$~\cite{Angloher:2015ewa}.  We
exemplify the near-future reach by following the experiment's own
projections~\cite{Angloher:2015eza} with threshold $100\,\eV$, a
factor of 100 reduced backgrounds, and neglecting efficiencies for
simplicity.

\paragraph{Comparison to DM-electron scattering.}

\begin{figure}
\centering
\includegraphics[width=\columnwidth]{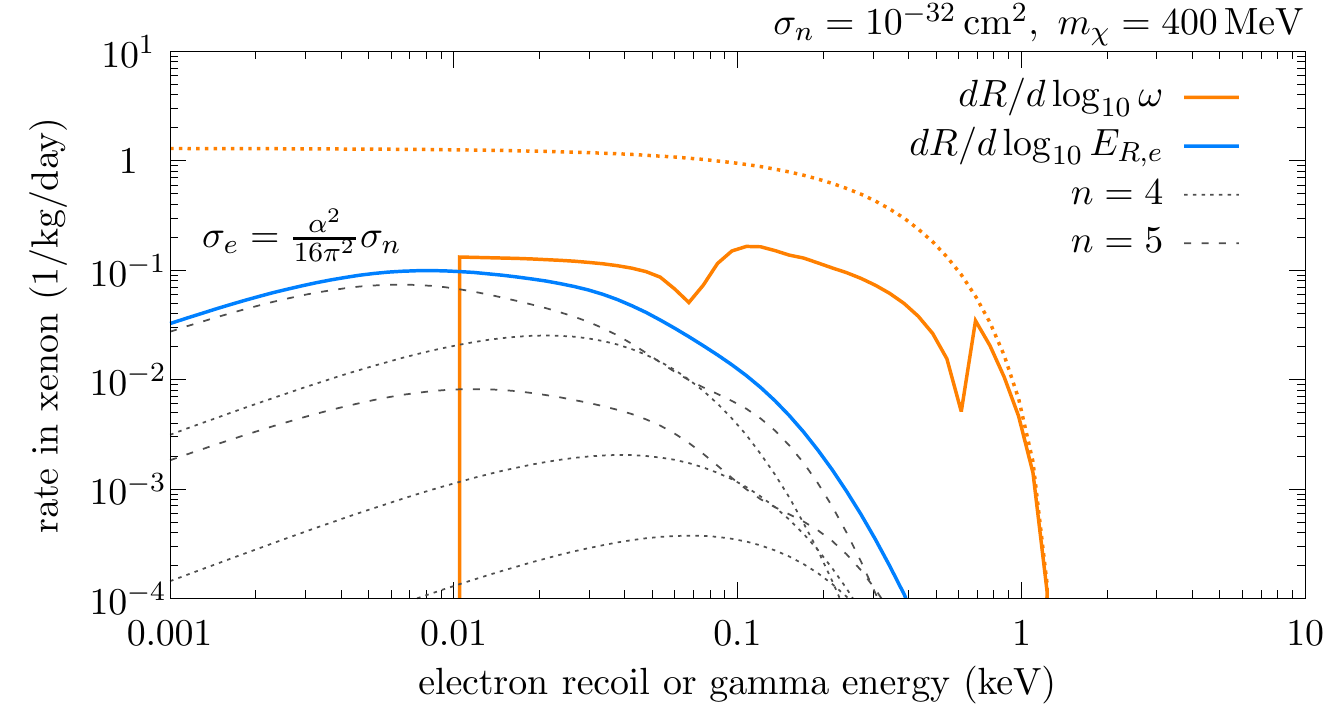}%
\caption{Comparison of electron recoil (blue) vs.~photon emission
  (orange) events in a leptophobic model  per logarithmic energy
  interval, $dR/d(\log_{10}E_{R,e})$ vs.~$dR/d(\log_{10}\omega)$,
  respectively, assuming that
  $\sigma_e = \alpha^2/(16\pi^2) \sigma_n$. The dotted (orange) line
  is the na\"ive rate. Thin gray lines break down the ionization
  contribution from the respective atomic shells (following previous calculations~\cite{Essig:2011nj,Essig:2012yx}.)}
\label{fig:compare}
\end{figure}

MeV-mass DM is already constrained through scattering on electrons and
the resulting ionization
signal~\cite{Essig:2011nj,Essig:2012yx}. 
Hence, best progress from our proposals can be expected in
``leptophobic'' models with suppressed (or absent) DM couplings to
electrons. %
One of the simplest leptophobic models is DM coupled to the SM through
a light $U(1)_B$ gauge boson $V_{\mu}$,
\begin{align}
\label{eq:L1}
{\cal L_{\rm int}} & = g_B (  V^\mu J^B_\mu  - \bar \chi \slashed V  \chi  ) - \frac{\kappa}{2} V^{\mu\nu} F_{\mu\nu} . 
\end{align}
with $g_B$ the $U(1)_B$ gauge coupling (charge),
$J_B^\mu \equiv \tfrac{1}{3} \sum_i \bar q_i \gamma^\mu q_i$ is the
baryon current (with the sum over all quark species).
Even if $\kappa = 0$ at the tree level, DM-electron scattering may be
induced radiatively, giving parametrically, if no cancellation occurs,
$\kappa_{\rm rad} \sim e g_B / (16\pi^2).$
This would lead to a ratio of cross sections of DM-electron over
DM-nucleon scattering as
$ {\sigma_{e}}/{\sigma_{n}} = {\alpha \kappa_{\rm rad}^2}/{\alpha_B}
\sim {\alpha^2}/{16\pi^2} \sim 3\times 10^{-7} $,
which demonstrates that a large hierarchy can be achieved in this
simple model;
$\sigma_{n} \simeq { 16\pi \alpha_B^2 m_{\chi}^2}{ m_V^{-4}}$ for
$m_V^2 \gg \vec q^2$.  This is exemplified in Fig.~\ref{fig:compare}
where we compare the rate of Bremsstrahlung emission to the rate of
DM-electron scattering $dR/dE_{R,e}$ for $m_{\chi} = 400\,\MeV$. Any
detailed analysis of electron multiplicity upon either scattering
process will likely improve the sensitivity to Bremsstrahlung because
of a higher primary energy of the photon.

The model of gauged baryon number is constrained in a number of ways,
notably from monojet production at colliders~\cite{Shoemaker:2011vi},
from missing energy contributions to rare meson
decays~\cite{Batell:2014yra}, and from cosmology; further,
model-dependent constraints arising from the UV-completion of $U(1)_B$
are obtained in~\cite{Dobrescu:2014fca}.
For illustration, in Fig.~\ref{fig:summary} we pick $m_V = 300\,\MeV$,
compatible with flavor constraints, and show the ensuing collider
limit which comes in this case from CDF. Finally, MiniBOONE may probe the window on large $\alpha_B$ and
MeV-scale $m_V$ in the near future~\cite{Batell:2014yra}.

\paragraph{Conclusions.}

In this letter we show that the irreducible contribution of photon
emission in the ordinary process of elastic DM-nucleus scattering, ``Bremsstrahlung,''
opens up the possibility to probe sub-GeV DM with present-day
technology and conventional detectors.
The photon endpoint energy is the kinetic energy of DM, 
and we derive the first limits on DM-nucleon scattering for
$m_{\chi}< 500\,\MeV$.

Further progress along the lines suggested here can be made. 
First, atomic physics calculations should allow to quantify
contributions to photon emission from excited final states of the
atom. 
Second, there is an additional contribution to the electron yield from
the ``shake-off'' of electrons in the elastic
scattering~\cite{Migdal:1977bq}.
Third, the spin-dependent case should be investigated. 
Fourth, signal formation in materials with band structure, where
single-atom polarizability is inadequate, should be investigated.
Fifth, photon emission in coherent neutrino nucleus scattering should
be included in direct detection neutrino background estimates.
A number of these points will be addressed in an upcoming
paper~\cite{CKJPupcoming}.

\paragraph{Acknowledgements} We are indebted to R.~Budnik, A.~G\"utlein,
F.~Kahlh\"ofer, J.~Mardon, D.~McKinsey, M.~Pospelov, M.~Pyle, and T.~Volansky for
useful discussions. CK is partially funded by the Danish National Research Foundation, grant number DNRF90. JP is supported by the New Frontiers program of
the Austrian Academy of Sciences.

\bibliography{biblio}

\begin{thebibliography}{48}%
\makeatletter
\providecommand \@ifxundefined [1]{%
 \@ifx{#1\undefined}
}%
\providecommand \@ifnum [1]{%
 \ifnum #1\expandafter \@firstoftwo
 \else \expandafter \@secondoftwo
 \fi
}%
\providecommand \@ifx [1]{%
 \ifx #1\expandafter \@firstoftwo
 \else \expandafter \@secondoftwo
 \fi
}%
\providecommand \natexlab [1]{#1}%
\providecommand \enquote  [1]{``#1''}%
\providecommand \bibnamefont  [1]{#1}%
\providecommand \bibfnamefont [1]{#1}%
\providecommand \citenamefont [1]{#1}%
\providecommand \href@noop [0]{\@secondoftwo}%
\providecommand \href [0]{\begingroup \@sanitize@url \@href}%
\providecommand \@href[1]{\@@startlink{#1}\@@href}%
\providecommand \@@href[1]{\endgroup#1\@@endlink}%
\providecommand \@sanitize@url [0]{\catcode `\\12\catcode `\$12\catcode
  `\&12\catcode `\#12\catcode `\^12\catcode `\_12\catcode `\%12\relax}%
\providecommand \@@startlink[1]{}%
\providecommand \@@endlink[0]{}%
\providecommand \url  [0]{\begingroup\@sanitize@url \@url }%
\providecommand \@url [1]{\endgroup\@href {#1}{\urlprefix }}%
\providecommand \urlprefix  [0]{URL }%
\providecommand \Eprint [0]{\href }%
\providecommand \doibase [0]{http://dx.doi.org/}%
\providecommand \selectlanguage [0]{\@gobble}%
\providecommand \bibinfo  [0]{\@secondoftwo}%
\providecommand \bibfield  [0]{\@secondoftwo}%
\providecommand \translation [1]{[#1]}%
\providecommand \BibitemOpen [0]{}%
\providecommand \bibitemStop [0]{}%
\providecommand \bibitemNoStop [0]{.\EOS\space}%
\providecommand \EOS [0]{\spacefactor3000\relax}%
\providecommand \BibitemShut  [1]{\csname bibitem#1\endcsname}%
\let\auto@bib@innerbib\@empty
\bibitem [{\citenamefont {Jungman}\ \emph {et~al.}(1996)\citenamefont
  {Jungman}, \citenamefont {Kamionkowski},\ and\ \citenamefont
  {Griest}}]{Jungman:1995df}%
  \BibitemOpen
  \bibfield  {author} {\bibinfo {author} {\bibfnamefont {G.}~\bibnamefont
  {Jungman}}, \bibinfo {author} {\bibfnamefont {M.}~\bibnamefont
  {Kamionkowski}}, \ and\ \bibinfo {author} {\bibfnamefont {K.}~\bibnamefont
  {Griest}},\ }\href {\doibase 10.1016/0370-1573(95)00058-5} {\bibfield
  {journal} {\bibinfo  {journal} {Phys. Rept.}\ }\textbf {\bibinfo {volume}
  {267}},\ \bibinfo {pages} {195} (\bibinfo {year} {1996})},\ \Eprint
  {http://arxiv.org/abs/hep-ph/9506380} {arXiv:hep-ph/9506380 [hep-ph]}
  \BibitemShut {NoStop}%
\bibitem [{\citenamefont {Bertone}\ \emph {et~al.}(2005)\citenamefont
  {Bertone}, \citenamefont {Hooper},\ and\ \citenamefont
  {Silk}}]{Bertone:2004pz}%
  \BibitemOpen
  \bibfield  {author} {\bibinfo {author} {\bibfnamefont {G.}~\bibnamefont
  {Bertone}}, \bibinfo {author} {\bibfnamefont {D.}~\bibnamefont {Hooper}}, \
  and\ \bibinfo {author} {\bibfnamefont {J.}~\bibnamefont {Silk}},\ }\href
  {\doibase 10.1016/j.physrep.2004.08.031} {\bibfield  {journal} {\bibinfo
  {journal} {Phys. Rept.}\ }\textbf {\bibinfo {volume} {405}},\ \bibinfo
  {pages} {279} (\bibinfo {year} {2005})},\ \Eprint
  {http://arxiv.org/abs/hep-ph/0404175} {arXiv:hep-ph/0404175 [hep-ph]}
  \BibitemShut {NoStop}%
\bibitem [{\citenamefont {Essig}\ \emph
  {et~al.}(2012{\natexlab{a}})\citenamefont {Essig}, \citenamefont {Mardon},\
  and\ \citenamefont {Volansky}}]{Essig:2011nj}%
  \BibitemOpen
  \bibfield  {author} {\bibinfo {author} {\bibfnamefont {R.}~\bibnamefont
  {Essig}}, \bibinfo {author} {\bibfnamefont {J.}~\bibnamefont {Mardon}}, \
  and\ \bibinfo {author} {\bibfnamefont {T.}~\bibnamefont {Volansky}},\ }\href
  {\doibase 10.1103/PhysRevD.85.076007} {\bibfield  {journal} {\bibinfo
  {journal} {Phys. Rev.}\ }\textbf {\bibinfo {volume} {D85}},\ \bibinfo {pages}
  {076007} (\bibinfo {year} {2012}{\natexlab{a}})},\ \Eprint
  {http://arxiv.org/abs/1108.5383} {arXiv:1108.5383 [hep-ph]} \BibitemShut
  {NoStop}%
\bibitem [{\citenamefont {Essig}\ \emph
  {et~al.}(2012{\natexlab{b}})\citenamefont {Essig}, \citenamefont
  {Manalaysay}, \citenamefont {Mardon}, \citenamefont {Sorensen},\ and\
  \citenamefont {Volansky}}]{Essig:2012yx}%
  \BibitemOpen
  \bibfield  {author} {\bibinfo {author} {\bibfnamefont {R.}~\bibnamefont
  {Essig}}, \bibinfo {author} {\bibfnamefont {A.}~\bibnamefont {Manalaysay}},
  \bibinfo {author} {\bibfnamefont {J.}~\bibnamefont {Mardon}}, \bibinfo
  {author} {\bibfnamefont {P.}~\bibnamefont {Sorensen}}, \ and\ \bibinfo
  {author} {\bibfnamefont {T.}~\bibnamefont {Volansky}},\ }\href {\doibase
  10.1103/PhysRevLett.109.021301} {\bibfield  {journal} {\bibinfo  {journal}
  {Phys. Rev. Lett.}\ }\textbf {\bibinfo {volume} {109}},\ \bibinfo {pages}
  {021301} (\bibinfo {year} {2012}{\natexlab{b}})},\ \Eprint
  {http://arxiv.org/abs/1206.2644} {arXiv:1206.2644 [astro-ph.CO]} \BibitemShut
  {NoStop}%
\bibitem [{\citenamefont {Graham}\ \emph {et~al.}(2012)\citenamefont {Graham},
  \citenamefont {Kaplan}, \citenamefont {Rajendran},\ and\ \citenamefont
  {Walters}}]{Graham:2012su}%
  \BibitemOpen
  \bibfield  {author} {\bibinfo {author} {\bibfnamefont {P.~W.}\ \bibnamefont
  {Graham}}, \bibinfo {author} {\bibfnamefont {D.~E.}\ \bibnamefont {Kaplan}},
  \bibinfo {author} {\bibfnamefont {S.}~\bibnamefont {Rajendran}}, \ and\
  \bibinfo {author} {\bibfnamefont {M.~T.}\ \bibnamefont {Walters}},\ }\href
  {\doibase 10.1016/j.dark.2012.09.001} {\bibfield  {journal} {\bibinfo
  {journal} {Phys. Dark Univ.}\ }\textbf {\bibinfo {volume} {1}},\ \bibinfo
  {pages} {32} (\bibinfo {year} {2012})},\ \Eprint
  {http://arxiv.org/abs/1203.2531} {arXiv:1203.2531 [hep-ph]} \BibitemShut
  {NoStop}%
\bibitem [{\citenamefont {Essig}\ \emph {et~al.}(2016)\citenamefont {Essig},
  \citenamefont {Fernandez-Serra}, \citenamefont {Mardon}, \citenamefont
  {Soto}, \citenamefont {Volansky},\ and\ \citenamefont {Yu}}]{Essig:2015cda}%
  \BibitemOpen
  \bibfield  {author} {\bibinfo {author} {\bibfnamefont {R.}~\bibnamefont
  {Essig}}, \bibinfo {author} {\bibfnamefont {M.}~\bibnamefont
  {Fernandez-Serra}}, \bibinfo {author} {\bibfnamefont {J.}~\bibnamefont
  {Mardon}}, \bibinfo {author} {\bibfnamefont {A.}~\bibnamefont {Soto}},
  \bibinfo {author} {\bibfnamefont {T.}~\bibnamefont {Volansky}}, \ and\
  \bibinfo {author} {\bibfnamefont {T.-T.}\ \bibnamefont {Yu}},\ }\href
  {\doibase 10.1007/JHEP05(2016)046} {\bibfield  {journal} {\bibinfo  {journal}
  {JHEP}\ }\textbf {\bibinfo {volume} {05}},\ \bibinfo {pages} {046} (\bibinfo
  {year} {2016})},\ \Eprint {http://arxiv.org/abs/1509.01598} {arXiv:1509.01598
  [hep-ph]} \BibitemShut {NoStop}%
\bibitem [{\citenamefont {Lee}\ \emph {et~al.}(2015)\citenamefont {Lee},
  \citenamefont {Lisanti}, \citenamefont {Mishra-Sharma},\ and\ \citenamefont
  {Safdi}}]{Lee:2015qva}%
  \BibitemOpen
  \bibfield  {author} {\bibinfo {author} {\bibfnamefont {S.~K.}\ \bibnamefont
  {Lee}}, \bibinfo {author} {\bibfnamefont {M.}~\bibnamefont {Lisanti}},
  \bibinfo {author} {\bibfnamefont {S.}~\bibnamefont {Mishra-Sharma}}, \ and\
  \bibinfo {author} {\bibfnamefont {B.~R.}\ \bibnamefont {Safdi}},\ }\href
  {\doibase 10.1103/PhysRevD.92.083517} {\bibfield  {journal} {\bibinfo
  {journal} {Phys. Rev.}\ }\textbf {\bibinfo {volume} {D92}},\ \bibinfo {pages}
  {083517} (\bibinfo {year} {2015})},\ \Eprint
  {http://arxiv.org/abs/1508.07361} {arXiv:1508.07361 [hep-ph]} \BibitemShut
  {NoStop}%
\bibitem [{\citenamefont {Guo}\ and\ \citenamefont
  {McKinsey}(2013)}]{Guo:2013dt}%
  \BibitemOpen
  \bibfield  {author} {\bibinfo {author} {\bibfnamefont {W.}~\bibnamefont
  {Guo}}\ and\ \bibinfo {author} {\bibfnamefont {D.~N.}\ \bibnamefont
  {McKinsey}},\ }\href {\doibase 10.1103/PhysRevD.87.115001} {\bibfield
  {journal} {\bibinfo  {journal} {Phys. Rev.}\ }\textbf {\bibinfo {volume}
  {D87}},\ \bibinfo {pages} {115001} (\bibinfo {year} {2013})},\ \Eprint
  {http://arxiv.org/abs/1302.0534} {arXiv:1302.0534 [astro-ph.IM]} \BibitemShut
  {NoStop}%
\bibitem [{\citenamefont {Schutz}\ and\ \citenamefont
  {Zurek}(2016)}]{Schutz:2016tid}%
  \BibitemOpen
  \bibfield  {author} {\bibinfo {author} {\bibfnamefont {K.}~\bibnamefont
  {Schutz}}\ and\ \bibinfo {author} {\bibfnamefont {K.~M.}\ \bibnamefont
  {Zurek}},\ }\href@noop {} {\  (\bibinfo {year} {2016})},\ \Eprint
  {http://arxiv.org/abs/1604.08206} {arXiv:1604.08206 [hep-ph]} \BibitemShut
  {NoStop}%
\bibitem [{\citenamefont {Hochberg}\ \emph
  {et~al.}(2016{\natexlab{a}})\citenamefont {Hochberg}, \citenamefont {Zhao},\
  and\ \citenamefont {Zurek}}]{Hochberg:2015pha}%
  \BibitemOpen
  \bibfield  {author} {\bibinfo {author} {\bibfnamefont {Y.}~\bibnamefont
  {Hochberg}}, \bibinfo {author} {\bibfnamefont {Y.}~\bibnamefont {Zhao}}, \
  and\ \bibinfo {author} {\bibfnamefont {K.~M.}\ \bibnamefont {Zurek}},\ }\href
  {\doibase 10.1103/PhysRevLett.116.011301} {\bibfield  {journal} {\bibinfo
  {journal} {Phys. Rev. Lett.}\ }\textbf {\bibinfo {volume} {116}},\ \bibinfo
  {pages} {011301} (\bibinfo {year} {2016}{\natexlab{a}})},\ \Eprint
  {http://arxiv.org/abs/1504.07237} {arXiv:1504.07237 [hep-ph]} \BibitemShut
  {NoStop}%
\bibitem [{\citenamefont {Hochberg}\ \emph {et~al.}(2015)\citenamefont
  {Hochberg}, \citenamefont {Pyle}, \citenamefont {Zhao},\ and\ \citenamefont
  {Zurek}}]{Hochberg:2015fth}%
  \BibitemOpen
  \bibfield  {author} {\bibinfo {author} {\bibfnamefont {Y.}~\bibnamefont
  {Hochberg}}, \bibinfo {author} {\bibfnamefont {M.}~\bibnamefont {Pyle}},
  \bibinfo {author} {\bibfnamefont {Y.}~\bibnamefont {Zhao}}, \ and\ \bibinfo
  {author} {\bibfnamefont {K.~M.}\ \bibnamefont {Zurek}},\ }\href@noop {} {\
  (\bibinfo {year} {2015})},\ \Eprint {http://arxiv.org/abs/1512.04533}
  {arXiv:1512.04533 [hep-ph]} \BibitemShut {NoStop}%
\bibitem [{\citenamefont {Cavoto}\ \emph {et~al.}(2016)\citenamefont {Cavoto},
  \citenamefont {Cirillo}, \citenamefont {Cocina}, \citenamefont {Ferretti},\
  and\ \citenamefont {Polosa}}]{Cavoto:2016lqo}%
  \BibitemOpen
  \bibfield  {author} {\bibinfo {author} {\bibfnamefont {G.}~\bibnamefont
  {Cavoto}}, \bibinfo {author} {\bibfnamefont {E.~N.~M.}\ \bibnamefont
  {Cirillo}}, \bibinfo {author} {\bibfnamefont {F.}~\bibnamefont {Cocina}},
  \bibinfo {author} {\bibfnamefont {J.}~\bibnamefont {Ferretti}}, \ and\
  \bibinfo {author} {\bibfnamefont {A.~D.}\ \bibnamefont {Polosa}},\ }\href
  {\doibase 10.1140/epjc/s10052-016-4193-7} {\bibfield  {journal} {\bibinfo
  {journal} {Eur. Phys. J.}\ }\textbf {\bibinfo {volume} {C76}},\ \bibinfo
  {pages} {349} (\bibinfo {year} {2016})},\ \Eprint
  {http://arxiv.org/abs/1602.03216} {arXiv:1602.03216 [physics.ins-det]}
  \BibitemShut {NoStop}%
\bibitem [{\citenamefont {Hochberg}\ \emph
  {et~al.}(2016{\natexlab{b}})\citenamefont {Hochberg}, \citenamefont {Kahn},
  \citenamefont {Lisanti}, \citenamefont {Tully},\ and\ \citenamefont
  {Zurek}}]{Hochberg:2016ntt}%
  \BibitemOpen
  \bibfield  {author} {\bibinfo {author} {\bibfnamefont {Y.}~\bibnamefont
  {Hochberg}}, \bibinfo {author} {\bibfnamefont {Y.}~\bibnamefont {Kahn}},
  \bibinfo {author} {\bibfnamefont {M.}~\bibnamefont {Lisanti}}, \bibinfo
  {author} {\bibfnamefont {C.~G.}\ \bibnamefont {Tully}}, \ and\ \bibinfo
  {author} {\bibfnamefont {K.~M.}\ \bibnamefont {Zurek}},\ }\href@noop {} {\
  (\bibinfo {year} {2016}{\natexlab{b}})},\ \Eprint
  {http://arxiv.org/abs/1606.08849} {arXiv:1606.08849 [hep-ph]} \BibitemShut
  {NoStop}%
\bibitem [{\citenamefont {Kouvaris}(2015)}]{Kouvaris:2015nsa}%
  \BibitemOpen
  \bibfield  {author} {\bibinfo {author} {\bibfnamefont {C.}~\bibnamefont
  {Kouvaris}},\ }\href {\doibase 10.1103/PhysRevD.92.075001} {\bibfield
  {journal} {\bibinfo  {journal} {Phys. Rev.}\ }\textbf {\bibinfo {volume}
  {D92}},\ \bibinfo {pages} {075001} (\bibinfo {year} {2015})},\ \Eprint
  {http://arxiv.org/abs/1506.04316} {arXiv:1506.04316 [hep-ph]} \BibitemShut
  {NoStop}%
\bibitem [{\citenamefont {Ellis}\ \emph {et~al.}(1988)\citenamefont {Ellis},
  \citenamefont {Flores},\ and\ \citenamefont {Lewin}}]{Ellis:1988nb}%
  \BibitemOpen
  \bibfield  {author} {\bibinfo {author} {\bibfnamefont {J.~R.}\ \bibnamefont
  {Ellis}}, \bibinfo {author} {\bibfnamefont {R.~A.}\ \bibnamefont {Flores}}, \
  and\ \bibinfo {author} {\bibfnamefont {J.~D.}\ \bibnamefont {Lewin}},\ }\href
  {\doibase 10.1016/0370-2693(88)91332-9} {\bibfield  {journal} {\bibinfo
  {journal} {Phys. Lett.}\ }\textbf {\bibinfo {volume} {B212}},\ \bibinfo
  {pages} {375} (\bibinfo {year} {1988})}\BibitemShut {NoStop}%
\bibitem [{\citenamefont {Baudis}\ \emph {et~al.}(2013)\citenamefont {Baudis},
  \citenamefont {Kessler}, \citenamefont {Klos}, \citenamefont {Lang},
  \citenamefont {Menéndez}, \citenamefont {Reichard},\ and\ \citenamefont
  {Schwenk}}]{Baudis:2013bba}%
  \BibitemOpen
  \bibfield  {author} {\bibinfo {author} {\bibfnamefont {L.}~\bibnamefont
  {Baudis}}, \bibinfo {author} {\bibfnamefont {G.}~\bibnamefont {Kessler}},
  \bibinfo {author} {\bibfnamefont {P.}~\bibnamefont {Klos}}, \bibinfo {author}
  {\bibfnamefont {R.~F.}\ \bibnamefont {Lang}}, \bibinfo {author}
  {\bibfnamefont {J.}~\bibnamefont {Menéndez}}, \bibinfo {author}
  {\bibfnamefont {S.}~\bibnamefont {Reichard}}, \ and\ \bibinfo {author}
  {\bibfnamefont {A.}~\bibnamefont {Schwenk}},\ }\href {\doibase
  10.1103/PhysRevD.88.115014} {\bibfield  {journal} {\bibinfo  {journal} {Phys.
  Rev.}\ }\textbf {\bibinfo {volume} {D88}},\ \bibinfo {pages} {115014}
  (\bibinfo {year} {2013})},\ \Eprint {http://arxiv.org/abs/1309.0825}
  {arXiv:1309.0825 [astro-ph.CO]} \BibitemShut {NoStop}%
\bibitem [{\citenamefont {McCabe}(2016)}]{McCabe:2015eia}%
  \BibitemOpen
  \bibfield  {author} {\bibinfo {author} {\bibfnamefont {C.}~\bibnamefont
  {McCabe}},\ }\href {\doibase 10.1088/1475-7516/2016/05/033} {\bibfield
  {journal} {\bibinfo  {journal} {JCAP}\ }\textbf {\bibinfo {volume} {1605}},\
  \bibinfo {pages} {033} (\bibinfo {year} {2016})},\ \Eprint
  {http://arxiv.org/abs/1512.00460} {arXiv:1512.00460 [hep-ph]} \BibitemShut
  {NoStop}%
\bibitem [{\citenamefont {Vergados}\ \emph {et~al.}(2016)\citenamefont
  {Vergados}, \citenamefont {Avignone}, \citenamefont {Kortelainen},
  \citenamefont {Pirinen}, \citenamefont {Srivastava}, \citenamefont
  {Suhonen},\ and\ \citenamefont {Thomas}}]{Vergados:2016ytt}%
  \BibitemOpen
  \bibfield  {author} {\bibinfo {author} {\bibfnamefont {J.~D.}\ \bibnamefont
  {Vergados}}, \bibinfo {author} {\bibfnamefont {F.~T.}\ \bibnamefont
  {Avignone}}, \bibinfo {author} {\bibfnamefont {M.}~\bibnamefont
  {Kortelainen}}, \bibinfo {author} {\bibfnamefont {P.}~\bibnamefont
  {Pirinen}}, \bibinfo {author} {\bibfnamefont {P.~C.}\ \bibnamefont
  {Srivastava}}, \bibinfo {author} {\bibfnamefont {J.}~\bibnamefont {Suhonen}},
  \ and\ \bibinfo {author} {\bibfnamefont {A.~W.}\ \bibnamefont {Thomas}},\
  }\href {\doibase 10.1088/0954-3899/43/11/115002} {\bibfield  {journal}
  {\bibinfo  {journal} {J. Phys.}\ }\textbf {\bibinfo {volume} {G43}},\
  \bibinfo {pages} {115002} (\bibinfo {year} {2016})},\ \Eprint
  {http://arxiv.org/abs/1601.06813} {arXiv:1601.06813 [hep-ph]} \BibitemShut
  {NoStop}%
\bibitem [{\citenamefont {Smith}\ \emph {et~al.}(2007)\citenamefont {Smith}
  \emph {et~al.}}]{Smith:2006ym}%
  \BibitemOpen
  \bibfield  {author} {\bibinfo {author} {\bibfnamefont {M.~C.}\ \bibnamefont
  {Smith}} \emph {et~al.},\ }\href {\doibase 10.1111/j.1365-2966.2007.11964.x}
  {\bibfield  {journal} {\bibinfo  {journal} {Mon. Not. Roy. Astron. Soc.}\
  }\textbf {\bibinfo {volume} {379}},\ \bibinfo {pages} {755} (\bibinfo {year}
  {2007})},\ \Eprint {http://arxiv.org/abs/astro-ph/0611671}
  {arXiv:astro-ph/0611671 [astro-ph]} \BibitemShut {NoStop}%
\bibitem [{\citenamefont {Akerib}\ \emph {et~al.}(2014)\citenamefont {Akerib}
  \emph {et~al.}}]{Akerib:2013tjd}%
  \BibitemOpen
  \bibfield  {author} {\bibinfo {author} {\bibfnamefont {D.~S.}\ \bibnamefont
  {Akerib}} \emph {et~al.} (\bibinfo {collaboration} {LUX}),\ }\href {\doibase
  10.1103/PhysRevLett.112.091303} {\bibfield  {journal} {\bibinfo  {journal}
  {Phys. Rev. Lett.}\ }\textbf {\bibinfo {volume} {112}},\ \bibinfo {pages}
  {091303} (\bibinfo {year} {2014})},\ \Eprint {http://arxiv.org/abs/1310.8214}
  {arXiv:1310.8214 [astro-ph.CO]} \BibitemShut {NoStop}%
\bibitem [{\citenamefont {Akerib}\ \emph
  {et~al.}(2016{\natexlab{a}})\citenamefont {Akerib} \emph
  {et~al.}}]{Akerib:2015rjg}%
  \BibitemOpen
  \bibfield  {author} {\bibinfo {author} {\bibfnamefont {D.~S.}\ \bibnamefont
  {Akerib}} \emph {et~al.} (\bibinfo {collaboration} {LUX}),\ }\href {\doibase
  10.1103/PhysRevLett.116.161301} {\bibfield  {journal} {\bibinfo  {journal}
  {Phys. Rev. Lett.}\ }\textbf {\bibinfo {volume} {116}},\ \bibinfo {pages}
  {161301} (\bibinfo {year} {2016}{\natexlab{a}})},\ \Eprint
  {http://arxiv.org/abs/1512.03506} {arXiv:1512.03506 [astro-ph.CO]}
  \BibitemShut {NoStop}%
\bibitem [{\citenamefont {Angloher}\ \emph {et~al.}(2016)\citenamefont
  {Angloher} \emph {et~al.}}]{Angloher:2015ewa}%
  \BibitemOpen
  \bibfield  {author} {\bibinfo {author} {\bibfnamefont {G.}~\bibnamefont
  {Angloher}} \emph {et~al.} (\bibinfo {collaboration} {CRESST}),\ }\href
  {\doibase 10.1140/epjc/s10052-016-3877-3} {\bibfield  {journal} {\bibinfo
  {journal} {Eur. Phys. J.}\ }\textbf {\bibinfo {volume} {C76}},\ \bibinfo
  {pages} {25} (\bibinfo {year} {2016})},\ \Eprint
  {http://arxiv.org/abs/1509.01515} {arXiv:1509.01515 [astro-ph.CO]}
  \BibitemShut {NoStop}%
\bibitem [{\citenamefont {Agnese}\ \emph {et~al.}(2016)\citenamefont {Agnese}
  \emph {et~al.}}]{Agnese:2015nto}%
  \BibitemOpen
  \bibfield  {author} {\bibinfo {author} {\bibfnamefont {R.}~\bibnamefont
  {Agnese}} \emph {et~al.} (\bibinfo {collaboration} {SuperCDMS}),\ }\href
  {\doibase 10.1103/PhysRevLett.116.071301} {\bibfield  {journal} {\bibinfo
  {journal} {Phys. Rev. Lett.}\ }\textbf {\bibinfo {volume} {116}},\ \bibinfo
  {pages} {071301} (\bibinfo {year} {2016})},\ \Eprint
  {http://arxiv.org/abs/1509.02448} {arXiv:1509.02448 [astro-ph.CO]}
  \BibitemShut {NoStop}%
\bibitem [{\citenamefont {Erickcek}\ \emph {et~al.}(2007)\citenamefont
  {Erickcek}, \citenamefont {Steinhardt}, \citenamefont {McCammon},\ and\
  \citenamefont {McGuire}}]{Erickcek:2007jv}%
  \BibitemOpen
  \bibfield  {author} {\bibinfo {author} {\bibfnamefont {A.~L.}\ \bibnamefont
  {Erickcek}}, \bibinfo {author} {\bibfnamefont {P.~J.}\ \bibnamefont
  {Steinhardt}}, \bibinfo {author} {\bibfnamefont {D.}~\bibnamefont
  {McCammon}}, \ and\ \bibinfo {author} {\bibfnamefont {P.~C.}\ \bibnamefont
  {McGuire}},\ }\href {\doibase 10.1103/PhysRevD.76.042007} {\bibfield
  {journal} {\bibinfo  {journal} {Phys. Rev.}\ }\textbf {\bibinfo {volume}
  {D76}},\ \bibinfo {pages} {042007} (\bibinfo {year} {2007})},\ \Eprint
  {http://arxiv.org/abs/0704.0794} {arXiv:0704.0794 [astro-ph]} \BibitemShut
  {NoStop}%
\bibitem [{\citenamefont {Batell}\ \emph {et~al.}(2014)\citenamefont {Batell},
  \citenamefont {deNiverville}, \citenamefont {McKeen}, \citenamefont
  {Pospelov},\ and\ \citenamefont {Ritz}}]{Batell:2014yra}%
  \BibitemOpen
  \bibfield  {author} {\bibinfo {author} {\bibfnamefont {B.}~\bibnamefont
  {Batell}}, \bibinfo {author} {\bibfnamefont {P.}~\bibnamefont
  {deNiverville}}, \bibinfo {author} {\bibfnamefont {D.}~\bibnamefont
  {McKeen}}, \bibinfo {author} {\bibfnamefont {M.}~\bibnamefont {Pospelov}}, \
  and\ \bibinfo {author} {\bibfnamefont {A.}~\bibnamefont {Ritz}},\ }\href
  {\doibase 10.1103/PhysRevD.90.115014} {\bibfield  {journal} {\bibinfo
  {journal} {Phys. Rev.}\ }\textbf {\bibinfo {volume} {D90}},\ \bibinfo {pages}
  {115014} (\bibinfo {year} {2014})},\ \Eprint {http://arxiv.org/abs/1405.7049}
  {arXiv:1405.7049 [hep-ph]} \BibitemShut {NoStop}%
\bibitem [{\citenamefont {Angle}\ \emph {et~al.}(2011)\citenamefont {Angle}
  \emph {et~al.}}]{Angle:2011th}%
  \BibitemOpen
  \bibfield  {author} {\bibinfo {author} {\bibfnamefont {J.}~\bibnamefont
  {Angle}} \emph {et~al.} (\bibinfo {collaboration} {XENON10}),\ }\href
  {\doibase 10.1103/PhysRevLett.110.249901, 10.1103/PhysRevLett.107.051301}
  {\bibfield  {journal} {\bibinfo  {journal} {Phys. Rev. Lett.}\ }\textbf
  {\bibinfo {volume} {107}},\ \bibinfo {pages} {051301} (\bibinfo {year}
  {2011})},\ \bibinfo {note} {[Erratum: Phys. Rev. Lett.110,249901(2013)]},\
  \Eprint {http://arxiv.org/abs/1104.3088} {arXiv:1104.3088 [astro-ph.CO]}
  \BibitemShut {NoStop}%
\bibitem [{\citenamefont {Aprile}\ \emph
  {et~al.}(2016{\natexlab{a}})\citenamefont {Aprile} \emph
  {et~al.}}]{Aprile:2016wwo}%
  \BibitemOpen
  \bibfield  {author} {\bibinfo {author} {\bibfnamefont {E.}~\bibnamefont
  {Aprile}} \emph {et~al.} (\bibinfo {collaboration} {XENON100}),\ }\href@noop
  {} {\  (\bibinfo {year} {2016}{\natexlab{a}})},\ \Eprint
  {http://arxiv.org/abs/1605.06262} {arXiv:1605.06262 [astro-ph.CO]}
  \BibitemShut {NoStop}%
\bibitem [{\citenamefont {Bloch}\ \emph {et~al.}(2016)\citenamefont {Bloch},
  \citenamefont {Essig}, \citenamefont {Tobioka}, \citenamefont {Volansky},\
  and\ \citenamefont {Yu}}]{Bloch:2016sjj}%
  \BibitemOpen
  \bibfield  {author} {\bibinfo {author} {\bibfnamefont {I.~M.}\ \bibnamefont
  {Bloch}}, \bibinfo {author} {\bibfnamefont {R.}~\bibnamefont {Essig}},
  \bibinfo {author} {\bibfnamefont {K.}~\bibnamefont {Tobioka}}, \bibinfo
  {author} {\bibfnamefont {T.}~\bibnamefont {Volansky}}, \ and\ \bibinfo
  {author} {\bibfnamefont {T.-T.}\ \bibnamefont {Yu}},\ }\href@noop {} {\
  (\bibinfo {year} {2016})},\ \Eprint {http://arxiv.org/abs/1608.02123}
  {arXiv:1608.02123 [hep-ph]} \BibitemShut {NoStop}%
\bibitem [{\citenamefont {Shutt}\ \emph {et~al.}(2007)\citenamefont {Shutt},
  \citenamefont {Dahl}, \citenamefont {Kwong}, \citenamefont {Bolozdynya},\
  and\ \citenamefont {Brusov}}]{Shutt:2006ed}%
  \BibitemOpen
  \bibfield  {author} {\bibinfo {author} {\bibfnamefont {T.}~\bibnamefont
  {Shutt}}, \bibinfo {author} {\bibfnamefont {C.~E.}\ \bibnamefont {Dahl}},
  \bibinfo {author} {\bibfnamefont {J.}~\bibnamefont {Kwong}}, \bibinfo
  {author} {\bibfnamefont {A.}~\bibnamefont {Bolozdynya}}, \ and\ \bibinfo
  {author} {\bibfnamefont {P.}~\bibnamefont {Brusov}},\ }\href {\doibase
  10.1016/j.nima.2007.04.104} {\bibfield  {journal} {\bibinfo  {journal} {Nucl.
  Instrum. Meth.}\ }\textbf {\bibinfo {volume} {A579}},\ \bibinfo {pages} {451}
  (\bibinfo {year} {2007})},\ \Eprint {http://arxiv.org/abs/astro-ph/0608137}
  {arXiv:astro-ph/0608137 [astro-ph]} \BibitemShut {NoStop}%
\bibitem [{\citenamefont {An}\ \emph {et~al.}(2013)\citenamefont {An},
  \citenamefont {Pospelov},\ and\ \citenamefont {Pradler}}]{An:2013yua}%
  \BibitemOpen
  \bibfield  {author} {\bibinfo {author} {\bibfnamefont {H.}~\bibnamefont
  {An}}, \bibinfo {author} {\bibfnamefont {M.}~\bibnamefont {Pospelov}}, \ and\
  \bibinfo {author} {\bibfnamefont {J.}~\bibnamefont {Pradler}},\ }\href
  {\doibase 10.1103/PhysRevLett.111.041302} {\bibfield  {journal} {\bibinfo
  {journal} {Phys. Rev. Lett.}\ }\textbf {\bibinfo {volume} {111}},\ \bibinfo
  {pages} {041302} (\bibinfo {year} {2013})},\ \Eprint
  {http://arxiv.org/abs/1304.3461} {arXiv:1304.3461 [hep-ph]} \BibitemShut
  {NoStop}%
\bibitem [{\citenamefont {An}\ \emph {et~al.}(2015)\citenamefont {An},
  \citenamefont {Pospelov}, \citenamefont {Pradler},\ and\ \citenamefont
  {Ritz}}]{An:2014twa}%
  \BibitemOpen
  \bibfield  {author} {\bibinfo {author} {\bibfnamefont {H.}~\bibnamefont
  {An}}, \bibinfo {author} {\bibfnamefont {M.}~\bibnamefont {Pospelov}},
  \bibinfo {author} {\bibfnamefont {J.}~\bibnamefont {Pradler}}, \ and\
  \bibinfo {author} {\bibfnamefont {A.}~\bibnamefont {Ritz}},\ }\href {\doibase
  10.1016/j.physletb.2015.06.018} {\bibfield  {journal} {\bibinfo  {journal}
  {Phys. Lett.}\ }\textbf {\bibinfo {volume} {B747}},\ \bibinfo {pages} {331}
  (\bibinfo {year} {2015})},\ \Eprint {http://arxiv.org/abs/1412.8378}
  {arXiv:1412.8378 [hep-ph]} \BibitemShut {NoStop}%
\bibitem [{\citenamefont {Huang}()}]{LUXtalk}%
  \BibitemOpen
  \bibfield  {author} {\bibinfo {author} {\bibfnamefont {D.}~\bibnamefont
  {Huang}},\ }\href@noop {} {\ }\Eprint {http://arxiv.org/abs/LUX
  collaboration. Talk given at UCLA Dark Matter 2016 workshop} {LUX
  collaboration. Talk given at UCLA Dark Matter 2016 workshop} \BibitemShut
  {NoStop}%
\bibitem [{\citenamefont {Akerib}\ \emph
  {et~al.}(2016{\natexlab{b}})\citenamefont {Akerib} \emph
  {et~al.}}]{Akerib:2015wdi}%
  \BibitemOpen
  \bibfield  {author} {\bibinfo {author} {\bibfnamefont {D.~S.}\ \bibnamefont
  {Akerib}} \emph {et~al.} (\bibinfo {collaboration} {LUX}),\ }\href {\doibase
  10.1103/PhysRevD.93.072009} {\bibfield  {journal} {\bibinfo  {journal} {Phys.
  Rev.}\ }\textbf {\bibinfo {volume} {D93}},\ \bibinfo {pages} {072009}
  (\bibinfo {year} {2016}{\natexlab{b}})},\ \Eprint
  {http://arxiv.org/abs/1512.03133} {arXiv:1512.03133 [physics.ins-det]}
  \BibitemShut {NoStop}%
\bibitem [{\citenamefont {Yellin}(2002)}]{Yellin:2002xd}%
  \BibitemOpen
  \bibfield  {author} {\bibinfo {author} {\bibfnamefont {S.}~\bibnamefont
  {Yellin}},\ }\href {\doibase 10.1103/PhysRevD.66.032005} {\bibfield
  {journal} {\bibinfo  {journal} {Phys. Rev.}\ }\textbf {\bibinfo {volume}
  {D66}},\ \bibinfo {pages} {032005} (\bibinfo {year} {2002})},\ \Eprint
  {http://arxiv.org/abs/physics/0203002} {arXiv:physics/0203002 [physics]}
  \BibitemShut {NoStop}%
\bibitem [{\citenamefont {McKinnsey}()}]{McKinnseyprivate}%
  \BibitemOpen
  \bibfield  {author} {\bibinfo {author} {\bibfnamefont {D.}~\bibnamefont
  {McKinnsey}},\ }\href@noop {} {\ }\Eprint {http://arxiv.org/abs/private
  communication} {private communication} \BibitemShut {NoStop}%
\bibitem [{\citenamefont {Collar}\ and\ \citenamefont
  {Avignone}(1992)}]{Collar:1992qc}%
  \BibitemOpen
  \bibfield  {author} {\bibinfo {author} {\bibfnamefont {J.~I.}\ \bibnamefont
  {Collar}}\ and\ \bibinfo {author} {\bibfnamefont {F.~T.}\ \bibnamefont
  {Avignone}},\ }\href {\doibase 10.1016/0370-2693(92)90873-3} {\bibfield
  {journal} {\bibinfo  {journal} {Phys. Lett.}\ }\textbf {\bibinfo {volume}
  {B275}},\ \bibinfo {pages} {181} (\bibinfo {year} {1992})}\BibitemShut
  {NoStop}%
\bibitem [{\citenamefont {Collar}\ and\ \citenamefont
  {Avignone}(1993)}]{Collar:1993ss}%
  \BibitemOpen
  \bibfield  {author} {\bibinfo {author} {\bibfnamefont {J.~I.}\ \bibnamefont
  {Collar}}\ and\ \bibinfo {author} {\bibfnamefont {F.~T.}\ \bibnamefont
  {Avignone}, \bibfnamefont {III}},\ }\href {\doibase 10.1103/PhysRevD.47.5238}
  {\bibfield  {journal} {\bibinfo  {journal} {Phys. Rev.}\ }\textbf {\bibinfo
  {volume} {D47}},\ \bibinfo {pages} {5238} (\bibinfo {year}
  {1993})}\BibitemShut {NoStop}%
\bibitem [{\citenamefont {Hasenbalg}\ \emph {et~al.}(1997)\citenamefont
  {Hasenbalg}, \citenamefont {Abriola}, \citenamefont {Avignone}, \citenamefont
  {Collar}, \citenamefont {Di~Gregorio}, \citenamefont {Gattone}, \citenamefont
  {Huck}, \citenamefont {Tomasi},\ and\ \citenamefont
  {Urteaga}}]{Hasenbalg:1997hs}%
  \BibitemOpen
  \bibfield  {author} {\bibinfo {author} {\bibfnamefont {F.}~\bibnamefont
  {Hasenbalg}}, \bibinfo {author} {\bibfnamefont {D.}~\bibnamefont {Abriola}},
  \bibinfo {author} {\bibfnamefont {F.~T.}\ \bibnamefont {Avignone}}, \bibinfo
  {author} {\bibfnamefont {J.~I.}\ \bibnamefont {Collar}}, \bibinfo {author}
  {\bibfnamefont {D.~E.}\ \bibnamefont {Di~Gregorio}}, \bibinfo {author}
  {\bibfnamefont {A.~O.}\ \bibnamefont {Gattone}}, \bibinfo {author}
  {\bibfnamefont {H.}~\bibnamefont {Huck}}, \bibinfo {author} {\bibfnamefont
  {D.}~\bibnamefont {Tomasi}}, \ and\ \bibinfo {author} {\bibfnamefont
  {I.}~\bibnamefont {Urteaga}},\ }\href {\doibase 10.1103/PhysRevD.55.7350}
  {\bibfield  {journal} {\bibinfo  {journal} {Phys. Rev.}\ }\textbf {\bibinfo
  {volume} {D55}},\ \bibinfo {pages} {7350} (\bibinfo {year} {1997})},\ \Eprint
  {http://arxiv.org/abs/astro-ph/9702165} {arXiv:astro-ph/9702165 [astro-ph]}
  \BibitemShut {NoStop}%
\bibitem [{\citenamefont {Zaharijas}\ and\ \citenamefont
  {Farrar}(2005)}]{Zaharijas:2004jv}%
  \BibitemOpen
  \bibfield  {author} {\bibinfo {author} {\bibfnamefont {G.}~\bibnamefont
  {Zaharijas}}\ and\ \bibinfo {author} {\bibfnamefont {G.~R.}\ \bibnamefont
  {Farrar}},\ }\href {\doibase 10.1103/PhysRevD.72.083502} {\bibfield
  {journal} {\bibinfo  {journal} {Phys. Rev.}\ }\textbf {\bibinfo {volume}
  {D72}},\ \bibinfo {pages} {083502} (\bibinfo {year} {2005})},\ \Eprint
  {http://arxiv.org/abs/astro-ph/0406531} {arXiv:astro-ph/0406531 [astro-ph]}
  \BibitemShut {NoStop}%
\bibitem [{\citenamefont {Aprile}\ \emph
  {et~al.}(2016{\natexlab{b}})\citenamefont {Aprile} \emph
  {et~al.}}]{Aprile:2015uzo}%
  \BibitemOpen
  \bibfield  {author} {\bibinfo {author} {\bibfnamefont {E.}~\bibnamefont
  {Aprile}} \emph {et~al.} (\bibinfo {collaboration} {XENON}),\ }\href
  {\doibase 10.1088/1475-7516/2016/04/027} {\bibfield  {journal} {\bibinfo
  {journal} {JCAP}\ }\textbf {\bibinfo {volume} {1604}},\ \bibinfo {pages}
  {027} (\bibinfo {year} {2016}{\natexlab{b}})},\ \Eprint
  {http://arxiv.org/abs/1512.07501} {arXiv:1512.07501 [physics.ins-det]}
  \BibitemShut {NoStop}%
\bibitem [{\citenamefont {Malling}\ \emph {et~al.}(2011)\citenamefont {Malling}
  \emph {et~al.}}]{Malling:2011va}%
  \BibitemOpen
  \bibfield  {author} {\bibinfo {author} {\bibfnamefont {D.~C.}\ \bibnamefont
  {Malling}} \emph {et~al.},\ }\href@noop {} {\  (\bibinfo {year} {2011})},\
  \Eprint {http://arxiv.org/abs/1110.0103} {arXiv:1110.0103 [astro-ph.IM]}
  \BibitemShut {NoStop}%
\bibitem [{\citenamefont {Pyle}()}]{MattPyleprivate}%
  \BibitemOpen
  \bibfield  {author} {\bibinfo {author} {\bibfnamefont {M.}~\bibnamefont
  {Pyle}},\ }\href@noop {} {\ }\Eprint {http://arxiv.org/abs/private
  communication} {private communication} \BibitemShut {NoStop}%
\bibitem [{\citenamefont {Derenzo}\ \emph {et~al.}(2016)\citenamefont
  {Derenzo}, \citenamefont {Essig}, \citenamefont {Massari}, \citenamefont
  {Soto},\ and\ \citenamefont {Yu}}]{Derenzo:2016fse}%
  \BibitemOpen
  \bibfield  {author} {\bibinfo {author} {\bibfnamefont {S.}~\bibnamefont
  {Derenzo}}, \bibinfo {author} {\bibfnamefont {R.}~\bibnamefont {Essig}},
  \bibinfo {author} {\bibfnamefont {A.}~\bibnamefont {Massari}}, \bibinfo
  {author} {\bibfnamefont {A.}~\bibnamefont {Soto}}, \ and\ \bibinfo {author}
  {\bibfnamefont {T.-T.}\ \bibnamefont {Yu}},\ }\href@noop {} {\  (\bibinfo
  {year} {2016})},\ \Eprint {http://arxiv.org/abs/1607.01009} {arXiv:1607.01009
  [hep-ph]} \BibitemShut {NoStop}%
\bibitem [{\citenamefont {Angloher}\ \emph {et~al.}(2015)\citenamefont
  {Angloher} \emph {et~al.}}]{Angloher:2015eza}%
  \BibitemOpen
  \bibfield  {author} {\bibinfo {author} {\bibfnamefont {G.}~\bibnamefont
  {Angloher}} \emph {et~al.} (\bibinfo {collaboration} {CRESST}),\ }\href@noop
  {} {\  (\bibinfo {year} {2015})},\ \Eprint {http://arxiv.org/abs/1503.08065}
  {arXiv:1503.08065 [astro-ph.IM]} \BibitemShut {NoStop}%
\bibitem [{\citenamefont {Shoemaker}\ and\ \citenamefont
  {Vecchi}(2012)}]{Shoemaker:2011vi}%
  \BibitemOpen
  \bibfield  {author} {\bibinfo {author} {\bibfnamefont {I.~M.}\ \bibnamefont
  {Shoemaker}}\ and\ \bibinfo {author} {\bibfnamefont {L.}~\bibnamefont
  {Vecchi}},\ }\href {\doibase 10.1103/PhysRevD.86.015023} {\bibfield
  {journal} {\bibinfo  {journal} {Phys. Rev.}\ }\textbf {\bibinfo {volume}
  {D86}},\ \bibinfo {pages} {015023} (\bibinfo {year} {2012})},\ \Eprint
  {http://arxiv.org/abs/1112.5457} {arXiv:1112.5457 [hep-ph]} \BibitemShut
  {NoStop}%
\bibitem [{\citenamefont {Dobrescu}\ and\ \citenamefont
  {Frugiuele}(2014)}]{Dobrescu:2014fca}%
  \BibitemOpen
  \bibfield  {author} {\bibinfo {author} {\bibfnamefont {B.~A.}\ \bibnamefont
  {Dobrescu}}\ and\ \bibinfo {author} {\bibfnamefont {C.}~\bibnamefont
  {Frugiuele}},\ }\href {\doibase 10.1103/PhysRevLett.113.061801} {\bibfield
  {journal} {\bibinfo  {journal} {Phys. Rev. Lett.}\ }\textbf {\bibinfo
  {volume} {113}},\ \bibinfo {pages} {061801} (\bibinfo {year} {2014})},\
  \Eprint {http://arxiv.org/abs/1404.3947} {arXiv:1404.3947 [hep-ph]}
  \BibitemShut {NoStop}%
\bibitem [{\citenamefont {Migdal}(1977)}]{Migdal:1977bq}%
  \BibitemOpen
  \bibfield  {author} {\bibinfo {author} {\bibfnamefont {A.~B.}\ \bibnamefont
  {Migdal}},\ }\href@noop {} {\bibfield  {journal} {\bibinfo  {journal} {Front.
  Phys.}\ }\textbf {\bibinfo {volume} {48}},\ \bibinfo {pages} {1} (\bibinfo
  {year} {1977})}\BibitemShut {NoStop}%
\bibitem [{\citenamefont {Kouvaris}\ and\ \citenamefont
  {Pradler}()}]{CKJPupcoming}%
  \BibitemOpen
  \bibfield  {author} {\bibinfo {author} {\bibfnamefont {C.}~\bibnamefont
  {Kouvaris}}\ and\ \bibinfo {author} {\bibfnamefont {J.}~\bibnamefont
  {Pradler}},\ }\href@noop {} {\ }\Eprint {http://arxiv.org/abs/in preparation}
  {in preparation} \BibitemShut {NoStop}%
\end{thebibliography}%
\end{document}